\documentclass[final,10pt,times]{elsarticle}

\usepackage{amssymb}
\usepackage{amsmath}
\usepackage{graphicx}
\usepackage{subcaption}
\usepackage{float}
\usepackage{hyperref}
\usepackage{subcaption}
\usepackage{xcolor}
\usepackage[numbers]{natbib}
\usepackage{geometry}
\journal{Elsevier}

\begin{document}

\begin{frontmatter}



\title{Constraining Fermionic Dark Matter with Galactic Neutron Stars}


\author[inst1,inst2]{Jianyuan Luo}
\author[inst3]{Dicong Liang\corref{cor1}}
\author[inst2,inst4]{Lijing Shao\corref{cor2}}

\affiliation[inst1]{organization={Department of Astronomy, School of Physic},
	addressline={Peking University}, 
	city={Beijing},
	postcode={100871}, 
	country={China}}

\affiliation[inst2]{organization={Kavli Institute for Astronomy and Astrophysics},
	addressline={Peking University}, 
	city={Beijing},
	postcode={100871}, 
	country={China}}

\affiliation[inst3]{organization={Department of Mathematics and Physics, School of Biomedical Engineering}, 
	addressline={Southern Medical University}, 
	city={Guangzhou},
	postcode={510515}, 
	country={China}}
	
\affiliation[inst4]{organization={National Astronomical Observatories},
            addressline={Chinese Academy of Sciences}, 
            city={Beijing},
            postcode={100012}, 
            country={China}}

\cortext[cor1]{Corresponding author: dcliang@smu.edu.cn }
\cortext[cor2]{Corresponding author: lshao@pku.edu.cn }

\begin{abstract}
Dark matter (DM) remains one of the most significant open questions in modern
physics, with its nature and interactions largely unexplored. In this study, we
investigate the behavior of massive fermionic DM particles in the context of
neutron stars (NSs), extending prior studies which focused on the bosonic DM. By
incorporating the  motion of NSs in the Galaxy and considering scenarios with
and without DM self-annihilation, we demonstrate their impact on the DM capture
rate and the accumulation process inside NSs. Observational data from pulsars in
the Milky Way are used to place constraints on DM properties, including the mass
and the DM-nucleon scattering cross-section, offering a more comprehensive
picture in probing DM interactions in astrophysical environments.
\end{abstract}

\end{frontmatter}


\section{Introduction}
\label{sec:1}

The concept of dark matter (DM) was first introduced when fast-moving galaxies
were observed in the Coma Cluster \cite{Zwicky:1933gu}.  Later, its existence
was further supported by the rotation curves of spiral galaxies
\cite{Rubin:1970zza}, gravitational lensing \cite{Clowe:2006eq}, the cosmic
microwave background (CMB) anisotropies \cite{Planck:2018vyg}, et cetera. 
Although DM does not interact electromagnetically, it participates in the
gravitational interaction.  While baryonic matter accounts for only about $5\%$
of the total mass of the Universe, DM is estimated to be five times more
abundant \cite{Planck:2018vyg}.  Despite its crucial role in the formation and
evolution of large-scale structure, the nature of DM remains poorly understood. 
Addressing the DM problem is essential for the advancement of particle physics
and astrophysics.

Besides various Earth-based probes \cite{XENON:2007uwm, XENON:2024wpa,
PandaX:2014mem, SuperCDMS:2020aus, Angloher:2011uu}, utilizing stellar objects
serves as a powerful approach in studying DM, offering a diverse range of DM
environments \cite{Linden:2024uph, Chakraborty:2024eyx, Nguyen:2025ygc, Adarsha:2025jqs, Bose:2023yll, Maity:2021fxw}. Compact stellar objects, such as neutron stars (NSs), have some
of the most extreme conditions, resulting in possible strong interactions with DM which can alter the stellar structure and evolution while producing detectable signals \cite{Bramante:2023djs, Baryakhtar:2017dbj, Gresham:2018rqo, East:2019dxt, Liu:2024swd, Bramante:2024ikc, Nguyen:2022zwb, Bell:2023ysh, Vikiaris:2023vau, Acevedo:2024ttq}.
Weakly interacting massive particles (WIMPs), a leading DM candidate
\cite{Roszkowski:2017nbc}, can be captured by NSs via scatterings with nucleons.
Once captured, WIMPs sink to the interior of the NS, reach thermal equilibrium
through continuous collisions, and ultimately accumulate near the core of the
star.  These particles may undergo gravitational collapse and form a small black
hole (BH) after reaching the Chandrasekhar limit. The BH can consume the star
and grow, leading to the destruction of the host star and finally leaving a
remnant BH with mass around 1 to 2 $M_\odot$ \cite{Dasgupta:2020mqg,
Garani:2021gvc}.  Therefore, observing intact NSs provides an opportunity to
place constraints on DM properties, such as the mass and the scattering cross
section with neutrons, which determine the possibility of NS destruction by
affecting the DM capture rate, collapse threshold, and so on.

While previous studies primarily treated NSs as static objects
\cite{Kouvaris:2007ay, Kouvaris:2010vv, Bramante:2014zca, Bell:2013xk}, Liang
and Shao \cite{Liang:2023nvo} provided a more realistic perspective and showed
that their motions in the Milky Way can significantly change the results. 
However, it only focused on bosonic particles, and the behaviours of fermions,
which are also possible candidates for WIMPs, remain unexplored in this context.
Also, Ref.~\cite{Liang:2023nvo} neglected DM self-annihilation, which has been
argued to result in sizable effects even for asymmetric DM
\cite{Bramante:2013nma, Bramante:2013hn, Petraki:2013wwa}.  Thus, this work
inherits the treatment of moving NSs and incorporates three key advancements:
(1) an investigation of fermionic DM and its comparison with bosons, (2) an
inclusion of DM self-annihilation, and (3) an extension beyond the NFW DM profile to
account for different distribution scenarios in the Galaxy. 

The main difference between fermions and bosons lies in the Chandrasekhar limit,
as fermions experience the degenerate pressure, which counteracts gravity and
suppresses BH formation.  The annihilation effect further reduces the likelihood
of DM collapse, and the distinction can only be quantified by numerically
calculating the accumulated particle number when considering the motion of NSs
in the Galaxy.  Additionally, the collapse likelihood depends on the DM density
profile, as it directly influences the capture rate.  By considering these
effects, we provide a more comprehensive examination of the NS destruction
caused by DM and leverage observations from pulsars to constrain DM properties,
thereby broadening the understanding of DM and its astrophysical implication.

The paper is organized as follows. In Sec.~\ref{sec:2}, we describe the process
that leads to the destruction of a NS, which is divided into five sub-processes.
Then in Sec.~\ref{sec:3}, we present the feasible collapse regions in the
parameter space of the DM particle mass $m_X$ and the DM-nucleon cross
section $\sigma_{nX}$, and compare fermions with bosons as well as
annihilating cases with non-annihilating cases.  Finally, we summarize our
results in Sec.~\ref{sec:4}.  Results from different density profiles are given
in the Appendix.

\section{Interaction of DM particles with NSs}
\label{sec:2}

In this section we briefly overview the five key sub-processes and the
corresponding physical parameters that underlie the scenario of NSs' accretion
of massive DM particles.

\subsection{Accumulation}

A DM particle is captured by a NS by scattering with neutrons on the NS surface
and losing energy. For simplicity, we use a toy model of NSs with a canonical mass
$M_{\mathrm {NS}} \simeq 1.5 \, M_\odot$, radius $R_{\mathrm {NS}} \simeq 10 \, {\mathrm {km}}$,
inner temperature $T_{\mathrm {NS}} \simeq 10^6  \,  {\mathrm K}$, and a uniform density. 
The escaping velocity at the surface of a NS is given by $v_{\mathrm {esc}}=\sqrt{2 G
M_{\mathrm {NS}} /R_{\mathrm {NS}}}$.
 We denote $m_X$ and $m_n$ the mass of DM particle and
neutron respectively, and denote $N_n$ the total number of neutrons within the
star. Taking relativistic effects into account, the capture rate is given by
\cite{Press:1985ug,Gould:1987ir,McDermott:2011jp}
\begin{equation}
F_{\mathrm {c}}=\sqrt{\frac{6}{\pi}}\frac{\rho_X}{\bar{v}_X}\frac{\xi N_n
v_{\mathrm {esc}}^2}{m_X}\left[1-\frac{1-\mathrm{exp}(-B^2)}{B^2}\right]\sigma_{\mathrm
{sat}}f(\sigma_{\mathrm{nX}}) \, \text{,}
\label{eq:1}
\end{equation}
where $\rho_X$ is the local DM density, and $\bar{v}_X$ is the DM velocity
dispersion.  The parameter, $\xi=\min \big\{m_X/(0.2\ {\mathrm {GeV}}),1 \big\}$,
accounts for the suppression in the capture rate due to the Pauli blocking
effect \cite{Bramante:2014zca,Bell:2013xk,McDermott:2011jp}. The parameter $B^2
= {6  \big(v_{\mathrm {esc}} / \bar{v}_X \big)^2m_Xm_n}/{ (m_X-m_n)^{2}}$,
describes the minimum energy loss necessary for the DM particle to be captured
\cite{McDermott:2011jp}, and the parameter, $f=1-\mathrm
{exp}\left(-\sigma_{\mathrm{nX}}/\sigma_{\mathrm {sat}}\right)$, describes the probability that
a DM particle scatters, where $\sigma_{\mathrm {sat}}=R_{\text{NS}}^2/(0.45\xi N_n)\simeq 10^{-45}\, \mathrm {cm}^2$. $f$ saturates to 1 at large cross sections $\sigma_{\mathrm{nX}}\gg\sigma_{\mathrm {sat}}$, showing the geometric limit of the capture rate \cite{Bramante:2013nma, Petraki:2013wwa}.

A single scattering is insufficient to capture DM particles with mass heavier than $10^6\,\mathrm{GeV}$ \cite{Baryakhtar:2017dbj, Joglekar:2019vzy, Joglekar:2020liw, Anzuini:2021lnv}. Therefore, multiple scatterings are required for these large-mass particles. However, when $\sigma_{\mathrm{nX}}\lesssim 10^{-45}\,\mathrm{cm^2}$ (i.e., when the NS is optically thin), multiple scatterings rarely occur, making the capture rate nearly identical to the single-scattering case \cite{Bell:2020jou, Bell:2020lmm, Bramante:2017xlb}. We therefore include multiple scatterings only for $m_X>10^6\,\mathrm{GeV}$ and $\sigma_{\mathrm{nX}}>10^{-45}\,\mathrm{cm^2}$. In this regime, the total capture rate is given by \cite{Bramante:2017xlb}
\begin{equation}
    F_{\mathrm{c}} = \sum_{i=1}^\infty  F_{\mathrm{c},i}\,\text{,}
\label{eq:2}
\end{equation}
where the capture rate after $i$ scatterings is
\begin{equation}
    F_{\mathrm{c},i} = \frac{1}{3} \sqrt{\frac{6}{\pi}}\frac{\rho_X}{\bar{v}_X}\frac{\pi R_{\mathrm{NS}}^2}{m_X} p_i(\tau)  
    \left[ 
        \big(2\bar{v}_X^2 + 3v_{\text{esc}}^2 \big) - 
        \big(2\bar{v}_X^2 + 3v_i^2 \big) \exp\left( -\frac{3\big(v_i^2 - v_{\text{esc}}^2\big)}{2\bar{v}_X^2} \right)
    \right]\,\text{,}
\label{eq:3}
\end{equation}
with $v_i = v_{\text{esc}} \left[1 - {2m_X m_n}/{(m_X + m_n)^2}\right]^{-i/2}$ being the critical velocity after $i$ scatterings, $p_i(\tau) = 2 \int_0^1 dy\,  {y e^{-y\tau} (y\tau)^i}/{i!}$ being the probability for $i$ scatterings, and $\tau = {3\sigma}/{(2\sigma_{\text{sat}})}$ being the optical depth. 
Notice that we have adopt a simplified model in this paper that the NS has a uniform density. 
For the more meticulous case where the (yet uncertain) NS inner structure is considered, one needs to solve the Tolman-Oppenheimer-Volkoff equations with a specific equation of state and the total capture rate needs to be calculated by integrating over the radius of NS. 
Besides, once taking into account the inner structure of a NS, the relativistic kinematics cannot be neglected any more \cite{Bell:2020obw}.
After incorporating the effects of nucleon structure and nucleon interactions, the capture rate can be suppressed substantially \cite{Bell:2020obw}. 
Interested readers may refer to Refs.~\cite{Anzuini:2021lnv, Bell:2020jou, Bell:2020lmm, Bell:2020obw} for more details.

As in Ref.~\cite{Gould:1987ir}, we let the DM particles have a Maxwell-Boltzmann distribution as seen by a moving observer,
then the capture rate is
modified as $F_{\mathrm c}(t)=\zeta F_{\mathrm c}$ after taking the Galactic motion of NSs into consideration.
Here $\zeta$ is the correction factor \cite{Gould:1987ir, Liang:2023nvo},
\begin{equation}
\begin{split}
\zeta &= \left[ 1 - \frac{1 - \exp(-B^2)}{B^2} \right]^{-1} 
\frac{(B_+ B_- - 1/2)[\chi(-\eta, \eta) - \chi(B_-, B_+) ]}{2 \eta B^2}\\
&+\frac{B_+ \exp(-B^2)/2 - B_- \exp(-B^2_+) /2 - \eta \exp(-\eta^2)}{2 \eta B^2}\,\text{,}\\
\end{split}
\label{eq:4}
\end{equation}
with $\chi(a, b) = \int_a^b \exp(-y^2)  dy = {\sqrt{\pi}} [\operatorname{erf}(b) - \operatorname{erf}(a)]/{2}$, $B_\pm = B \pm \eta$ and $\eta = \sqrt{3/2}  v_{\star} / \bar{v}_X$, where $v_{\star}$ is the relative velocity between the NS and the DM halo. In our investigation, both $v_{\star}$ and $\bar{v}_X$ are time-dependent.

\begin{figure}[t] 
            \centering
            \includegraphics[width=0.5\linewidth]{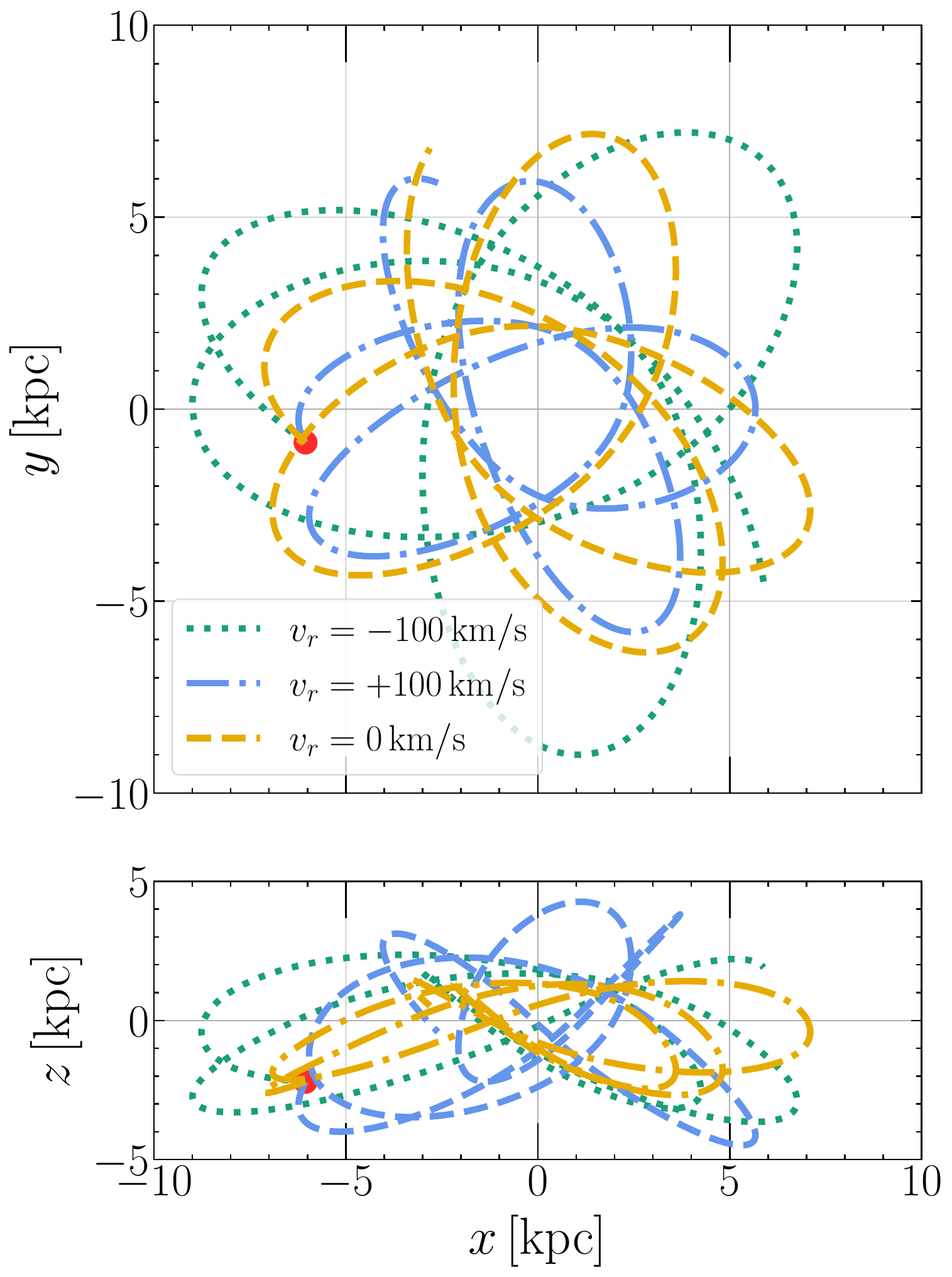} 
        \caption{Trajectory of PSR J2129$-$5721 in the past 500\,Myr. The red
        dot represents its current position. The radial velocity $v_r$ is set
        manually and three cases are considered where $v_r =
        -100\,\mathrm{km/s}$, $0\,\mathrm{km/s}$ and $100\,\mathrm{km/s}$.}
        \label{fig:1}
\end{figure}

The DM density $\rho_X$ is also time-dependent and 
determined by the location of the NS and the DM density
profile of the Galaxy, which we adopt the Navarro-Frenk-White profile
\cite{Navarro:1996gj}
\begin{equation}
    \rho_X(r)=\frac{\rho_{0}}{(r/r_0)^{\gamma}(1+r/r_0)^{3-\gamma}} \,,
    \label{eq:5}
\end{equation}
where $r_0=18.6 \, {\mathrm {kpc}}$ and $\rho_{0}$ is a normalization factor such that
the density at the Solar system is $\rho_{X} \big(r=8.20 \, {\mathrm {kpc}} \big)
=0.38 \, {\mathrm {GeV}}/{\mathrm {cm}}^3$.  Here we adopt $\gamma=1$, while other
possibilities are considered in the Appendix.  As the DM environment changes,
the potential in the Galaxy changes correspondingly.  Exploiting current
position and velocity information of the pulsars from the Australia Telescope
National Facility (ATNF) pulsar catalogue \cite{Manchester:2004bp}, the
trajectories in the past can be inferred with the {\tt GalPot} package\footnote{
\url{https://github.com/PaulMcMillan-Astro/GalPot/tree/master}}
\cite{2016ascl.soft11006M,McMillan:2016jtx} and thus we get $\rho_X(t)$, $\bar{v}_X(t)$ and $\zeta(t)$.  Due to the limitations of
observation, the radial velocity of the pulsar, $v_r$, is set manually.  As in
Ref.~\cite{Liang:2023nvo}, we consider three cases, $v_r= -100\,\mathrm{km/s}$,
$v_r=0\,\mathrm{km/s}$, and $v_r= 100\,\mathrm{km/s}$.  For example, the orbit of PSR
J2129$-$5721 in the past 500\,Myr is shown in Fig.~\ref{fig:1}. 

\subsection{Thermalization}
After being captured by a NS, DM particles continue to undergo scatterings with neutrons and
further lose energy until thermal equilibrium is reached. This process is
called thermalization. Using the differential scattering rate, we can derive the energy loss per unit time and thus estimate the thermalization timescale as \cite{Bell:2023ysh}
\begin{equation}
  t_{\mathrm {th}}=\frac{147\pi^2}{16}\frac{m_X}{(m_n^{\text{eff}}+m_X)^2}\frac{1}{\sigma_{\mathrm{nX}}}\frac{1}{(k_bT_{\mathrm{NS}})^2} \,,
  \label{eq:6}
\end{equation}
where $m_n^{\text{eff}}=0.5\,m_n$ is adopted as the target mass. For large mass $m_X\gg m_n$, it can be approximated as \cite{Bell:2023ysh}
\begin{equation}
  t_{\mathrm {th}}=9.7\times10^1\,\mathrm{yrs}\left(\frac{\mathrm{GeV}}{m_X}\right)\left(\frac{10^{-45}\,\mathrm{cm^2}}{\sigma_{\mathrm{nX}}}\right)\left(\frac{10^{6}\,\mathrm{K}}{T_{\mathrm{NS}}}\right)^2\,.
  \label{eq:7}
\end{equation}

The volume occupied by thermalized DM particles is evaluated using the virial
theorem
\begin{equation}
	\frac{3}{2}k_bT_{\mathrm {NS}}=\frac{2\pi G\rho_nm_Xr^2}{3} \,\text{,}
	\label{eq:8}
\end{equation}
where $\rho_n$ is the number density of the neutron, and $r$ is the thermal
radius
\begin{equation}
	r_{\mathrm{th}}=\sqrt{\frac{9k_bT_{\mathrm {NS}}}{4\pi G\rho_nm_X}} \, .
	\label{eq:9}
\end{equation}
Thus, we get the volume of the thermalized DM core,
\begin{equation}
	\begin{aligned}
		V_{\mathrm{th}}=\frac{4}{3}\pi r_{\mathrm{th}}^3=\frac{4}{3}\pi
		\left(\frac{9k_bT_{\mathrm {NS}}}{4\pi G\rho_nm_X} \right)^{3/2}\text{.}
	\end{aligned}
	\label{eq:10}
\end{equation}

\subsection{Annihilation}

The number of accumulated DM particles will increase with time but it can be
diminished if we consider self-annihilation at the same time.  The annihilation
rate was estimated as \cite{Bramante:2013nma}
\begin{equation}
	F_{\mathrm a}=-\frac{\langle\sigma_{\mathrm a}v\rangle N_X^2}{V_{\mathrm {th}}} \, \text{,}
	\label{eq:11}
\end{equation}
where $\langle\sigma_{\mathrm a}v\rangle$ is the effective annihilation  cross section. Following the range used in Ref.~\cite{Bramante:2013hn}, we adopted $10^{-48}\,\mathrm{cm^3/s}$ as an optimal value: smaller cross sections render the annihilation effect insignificant, while larger values weaken the constraints on DM particles due to excessive annihilation. Therefore, the rate at which
DM accumulates inside the star can be obtained by solving the following
differential equation
\begin{equation}
	\frac{{\mathrm d} N_X}{{\mathrm d} t}=F_{\mathrm c}-\frac{\langle\sigma_{\mathrm a}v\rangle
	N_X^2}{V_{\mathrm {th}}} \, \text{.}
	\label{eq:12}
\end{equation}

\subsection{Collapse}

When $N_X$ increases to a certain extent, the gravitational interaction of the
DM becomes dominant. This happens when the total mass of DM becomes greater than
the total mass of neutrons within the thermalized volume $V_{\mathrm {th}}$
\cite{McDermott:2011jp}, 
\begin{equation}
N_Xm_X>\frac{4\pi}{3}\rho_nr_{\mathrm{th}}^3 \, \text{,}
\label{eq:13}
\end{equation}
and the corresponding DM number is 
\begin{equation}
N_{\mathrm{self}} = \frac{4\pi}{3} \frac{\rho_nr_{\mathrm{th}}^3}{m_X} \,
\text{.}
\label{eq:14}
\end{equation}
DM particles begin to form a self-gravitating system when
$N_X>N_{\mathrm{self}}$, and if the number exceeds Chandrasekhar limit
simultaneously, the DM starts to collapse.  Specifically, the Chandrasekhar
limit for fermionic matter is given by \cite{Shapiro:1983du}
\begin{equation}
N_{\mathrm{chan, fer}}= \frac{m_{\mathrm{pl}}^3}{g_f^{1/2}m_X^3} \, \text{.}
\label{eq:15}
\end{equation}
Therefore, the collapse threshold is  
\begin{equation}
N_{\mathrm{coll}}=\mathrm{max}\, \Big[N_{\mathrm{self}}, \, N_{\mathrm{chan,
fer}} \Big] \, \text{.}
\label{eq:16}
\end{equation}

\subsection{BH growth}

After the DM core collapses into a BH, its mass should increase over time to be
long-lasting so that it can destroy the host star.  Main contributions to the
change of BH mass include neutron accretion, DM accretion and Hawking radiation
\cite{Hawking:1974rv}.  For fermions, DM accretion is negligible, so the
evolution of the BH mass, $M_{\mathrm {BH}}$, follows \cite{Bramante:2013nma,
Bondi:1952ni}
\begin{equation}
\frac{{\mathrm d} M_{\mathrm {BH}}}{{\mathrm d} t}=\frac{4\pi\lambda G^2 M_{\mathrm
{BH}}^2\rho_n}{v_s^3}-\frac{1}{15360\pi G^2M_{\mathrm {BH}}^2} \, \text{.}
\label{eq:17}
\end{equation}
Here, $\lambda_s=0.25$ is the accretion eigenvalue for the transonic solution,
and $v_s=0.33 \, c$ is the sound speed in the NS.

\section{Results and Discussions}
\label{sec:3}

Given the above scenario, a BH formed from DM particles inside the NS will
destroy the host star if the following three conditions are satisfied
simultaneously, (i)  Thermalization Condition: $t_{\mathrm {th}}<t_{\mathrm {NS}}$; (ii)
Collapse Condition: $N_X>N_{\mathrm {coll}}$; and (iii) Growth Condition: ${{\mathrm d}
M_{\mathrm {BH}}} / {{\mathrm d} t}>0$.

Using Eqs.~\eqref{eq:6} and ~\eqref{eq:7}, we have 
\begin{equation}
\sigma_{\mathrm{nX}}>10^{-45}\,\mathrm{cm^2}
\left(\frac{2.1\times10^2\,\mathrm{yrs}}{t_{\mathrm{NS}}}\right)
\frac{m_X/m_n^{\text{eff}}}{\big(1+m_X/m_n^{\text{eff}} \big)^2}\left(\frac{10^6\,\mathrm{K}}{T_{\mathrm
{NS}}}\right)^2 \,\text{,}
\label{eq:18}
\end{equation}
for the Thermalization Condition and 
\begin{equation}
\sigma_{\mathrm{nX}}>10^{-45}\,\mathrm{cm^2}
\left(\frac{9.7\times10^1\,\mathrm{yrs}}{t_{\mathrm{NS}}}\right)
\left(\frac{\mathrm{GeV}}{m_X}\right) \left(\frac{10^6\,\mathrm{K}}{T_{\mathrm
{NS}}}\right)^2 \,\text{,}
\label{eq:19}
\end{equation}
at large mass.

Using Eq.~\eqref{eq:17} and $M_{\mathrm{BH}}=N_{\mathrm {coll}}m_{X}$, we have 
\begin{equation}
N_{\mathrm{coll}}m_X>\frac{1}{G}\left( \frac{v_s^3}{61440\pi^2\lambda\rho_n}
\right)^{1/4} \, \text{.}
\label{eq:20} 
\end{equation}
Plugging in $N_{\mathrm{coll}}=N_{\mathrm{chan, fer}}$, we get the upper limit
\begin{equation}
m_X\lesssim10^{10}\,\mathrm{GeV} \, \text{,}
\label{eq:21}
\end{equation}
for the Growth Condition. 

\begin{figure}[t] 
            \centering
            \includegraphics[width=0.5\linewidth]{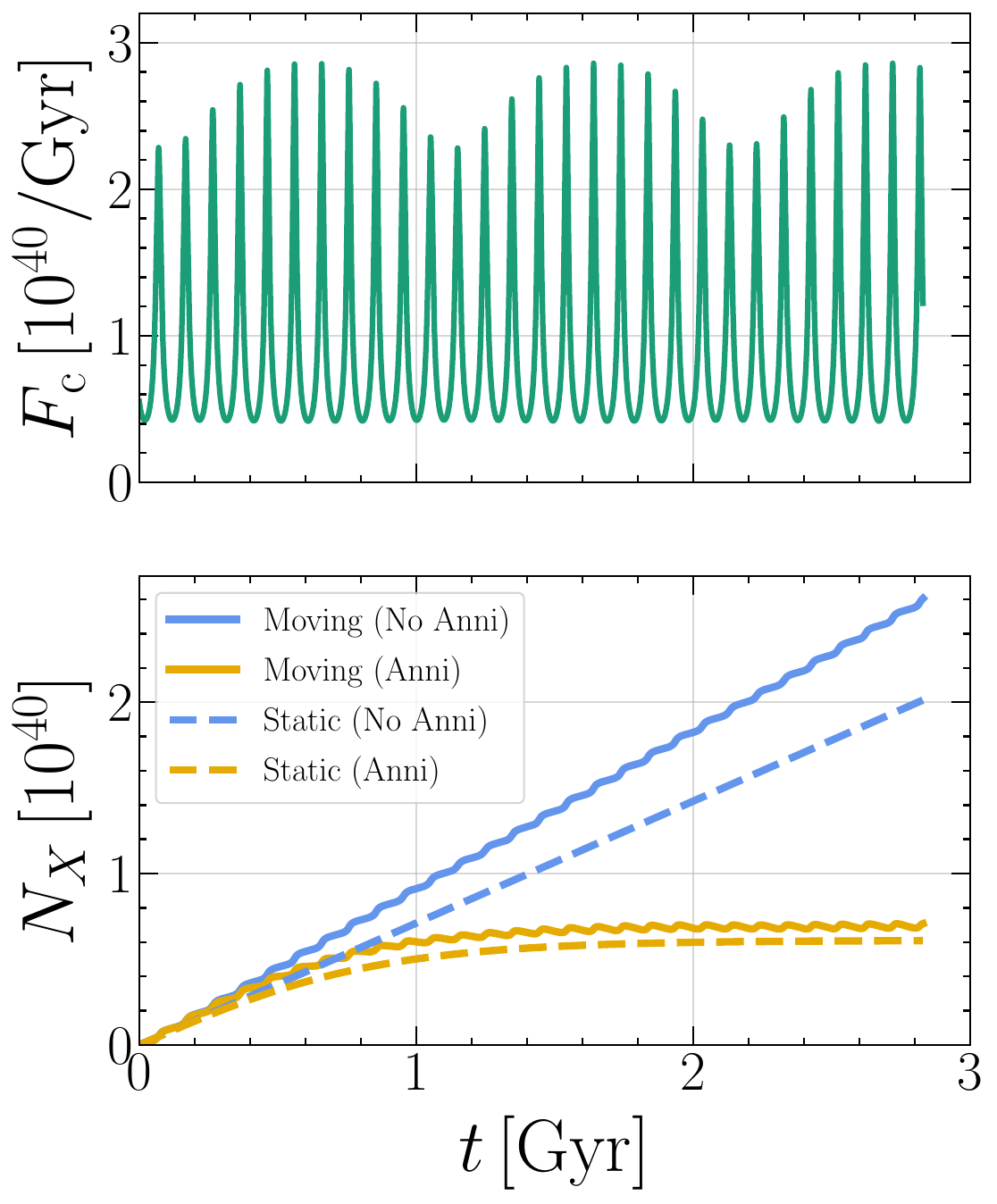} 
        \caption{The capture rate $F_{\mathrm c}$ \textit{(top)} and the accumulated particle
        number $N_X$ \textit{(bottom)} as functions of time for PSR J2129$-$5721, assuming $R_{\mathrm
        NS}=10\, \mathrm {km}$, $M_{\mathrm {NS}}=1.5\, M_{\odot}$,  $T_{\mathrm {NS}}=10^6\,\mathrm
        K$, and $\langle\sigma_{\mathrm{a}}v\rangle=10^{-48}\,\mathrm{cm^3/s}$. In the bottom panel, blue lines correspond to non-annihilation cases, while yellow lines show annihilation cases; solid lines incorporate NS motion, while dashed lines assume a static NS. $F_{\mathrm
        c}$ changes periodically and $N(t)$ increases wavily as the NS orbits
        around the Galaxy. Parameters $m_X=10^1\,\mathrm{GeV}$,
        $\sigma_{\mathrm{nX}}=10^{-46}\,\mathrm{cm^2}$, and $v_r=0\,\mathrm{km/s}$  are
        used in this example.}
        \label{fig:2}
\end{figure}

According to Eq.~\eqref{eq:12}, if we assume the NS is static, the number of
accumulated DM particles $N_X$ can be solved as 
\begin{equation}
N_X=F_{\mathrm c}t
\label{eq:22}
\end{equation}
and 
\begin{equation}
N_X=\sqrt{\frac{F_{\mathrm c}V_{\mathrm {th}}}{\langle\sigma_{\mathrm{a}}v\rangle}} \, \mathrm
{tanh} \left(\sqrt{\frac{F_{\mathrm {c}}\langle\sigma_{\mathrm{a}}v\rangle}{V_{\mathrm
{th}}}}t\right) \, \text{,}
\label{eq:23}
\end{equation}
for the cases of non-annihilation and annihilation, respectively.  We see that
in the case of non-annihilation, $N_X$ increases continuously, while in the case
of annihilation, due to the property of the $\mathrm {tanh}$ function, $N_X$ reaches
an asymptotic value 
\begin{equation}
N_{\mathrm {asy}}=\sqrt{\frac{F_{\mathrm c}V_{\mathrm {th}}}{\langle\sigma_{\mathrm{a}}v\rangle}} \,,
\label{eq:24}
\end{equation}
around 
\begin{equation}
t_{\mathrm {asy}}=2\sqrt{\frac{V_{\mathrm {th}}}{F_{\mathrm {c}}\langle\sigma_{\mathrm{a}}v\rangle}} \,,
\label{eq:25}
\end{equation}
after which the increment becomes negligible. If a NS's motion is considered,
the capture rate $F_{\mathrm c}$ becomes periodical and $N(t)$ increases wavily as
the NS orbits around the center of the Galaxy, which is shown for PSR
J2129$-$5721 in Fig.~\ref{fig:2}.

\begin{figure}[t] 
            \centering
            \includegraphics[width=0.9\linewidth]{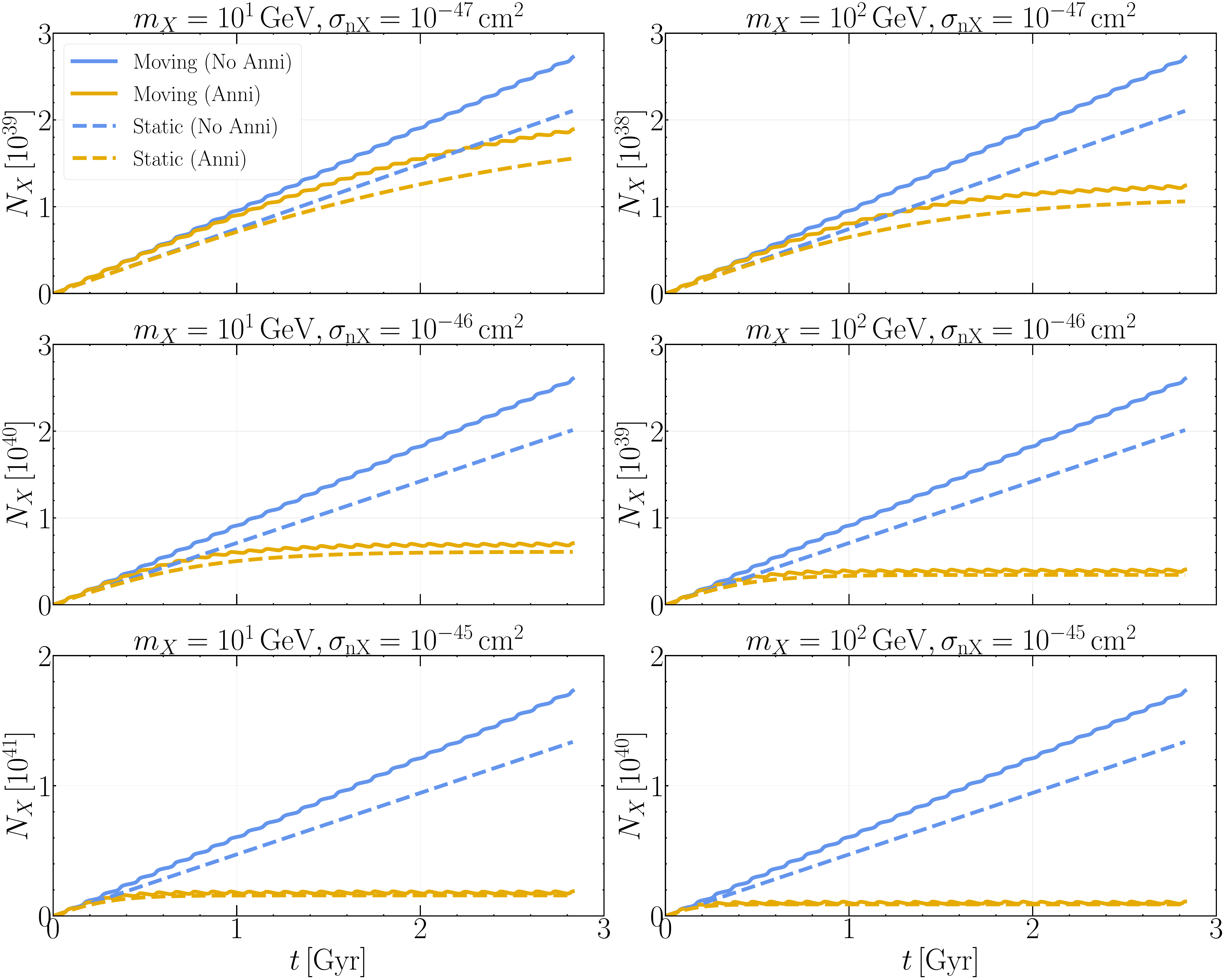} 
        \caption{Time evolutions of the accumulated particle number $N_X$ for
        PSR J2129$-$5721. The same settings of parameters are used as in
        Fig.~\ref{fig:2}. 
        }
        \label{fig:3}
\end{figure}

From Eq.~\eqref{eq:1}, when $m_X \ll 10^6 \, \mathrm{GeV}$ and $\sigma_{\mathrm{nX}}\ll 10^{-45}\,\mathrm{cm^2}$, we have $1- \big({1- e^{-B^2}}\big)/{B^2}\approx1$ and $1 - e^{-\sigma_{\mathrm{nX}}/\sigma_{\mathrm
{sat}}} \approx \sigma_{\mathrm{nX}}$. Therefore, 
\begin{equation}
F_c \sim \frac{\left[1-\frac{1- e^{-B^2}}{B^2}\right]}{m_X} \left(1 - e^{-\sigma_{\mathrm{nX}}/\sigma_{\mathrm
{sat}}}\right) \sim \frac{\sigma_{\mathrm{nX}}}{m_X}\,.
\label{eq:26}
\end{equation}
From Eq.~\eqref{eq:10}, we have $V_{\mathrm{th}}\sim {m_X^{-3/2}}$. A a result, we get
\begin{equation}
    N_{\mathrm{asy}}\sim {\sigma_{\mathrm{nX}}^{1/2}}{m_X^{-5/4}}\,,
\label{eq:27}
\end{equation}
and 
\begin{equation}
t_{\mathrm{asy}}\sim\sigma_{\mathrm{nX}}^{-1/2}m_X^{-1/4}\,. 
\label{eq:28}
\end{equation}
Thus, when $m_X$ increases, $N_{\mathrm {asy}}$ and $t_{\mathrm {asy}}$ both decrease. Also, when $\sigma_{\mathrm{nX}}$ increases, $N_{\mathrm {asy}}$ increases and $t_{\mathrm {asy}}$ decreases, as shown in Fig.~\ref{fig:3}. Even if $\sigma_{\mathrm{nx}}$ approaches $10^{-45}\,\text{GeV}$, the trend remains the same because $1 - \exp(-\sigma_{\mathrm{nX}}/\sigma_{\mathrm
{sat}})$ is still increasing with $\sigma_{\mathrm{nX}}$. These are what we would expect intuitively, since DM particles slow down with a heavier mass, leading to fewer particles being captured. A larger cross section means a stronger interaction, so that more particles are deposited.  These trends are the same if NS motions are considered, because the modification in Eq.~\eqref{eq:4} varies with time periodically, and it is independent of $m_X$ or $\sigma_{\mathrm{nX}}$.
When $m_X \gg 10^6 \,\mathrm{GeV}$, we have $\left[1- \big({1-e^{-B^2}}\big)/{B^2}\right] \sim {1}/{m_X}$, which leads to $N_{\mathrm{asy}} \sim m_X^{-7/4}$. 
It remains valid that NSs capture fewer particles as the DM mass increases.


In the case of non-annihilation and single scattering, $N_X=\int_0^{t_{\mathrm {NS}}} F_{\mathrm c}(t) \, {\mathrm
d}t$, and the Collapse Condition  becomes
\begin{equation}
\sigma_{\mathrm{nX}}>\sigma_{\mathrm{sat}}\ln \left( 1 -
\frac{N_{\mathrm{coll}}m_X}{PAt_{\mathrm{NS}}} \right)^{-1} \,\text{,}
\label{eq:29}
\end{equation}
where 
\begin{equation}
  P=\sqrt{\frac{6}{\pi}}\frac{\xi N_n v_{\mathrm {esc}}^2\sigma_{\mathrm {sat}}}{v_0} \,
  \text{,}
\label{eq:30}
\end{equation}
and
\begin{equation}
  A=\frac{1}{t_{\mathrm {NS}}}\int_0^{t_{\mathrm
  {NS}}}\zeta(t)\rho_X(t)\frac{v_0}{\bar{v}(t)} {\mathrm d} t \, \text{,}
\label{eq:31}
\end{equation}
with $v_0=200\,\mathrm{km/s}$. In the case of annihilation or multiple scattering, numerical solution is needed.  We evenly take
points within the region of interest in the $m_X$-$\sigma_{\mathrm{nX}}$ parameter space.
For each point, we use the forward Euler method to solve Eq.~\eqref{eq:12} with
the initial condition $N_X(t=0)=0$. Then, we compare $N_X$ with $N_{\mathrm {coll}}$
point by point and label the ones where $N_X/N_{\mathrm {coll}}>1$. Thus, we get the
feasible parameter space for the Collapse Condition.

\begin{figure}[t] 
            \centering
            \includegraphics[width=\linewidth]{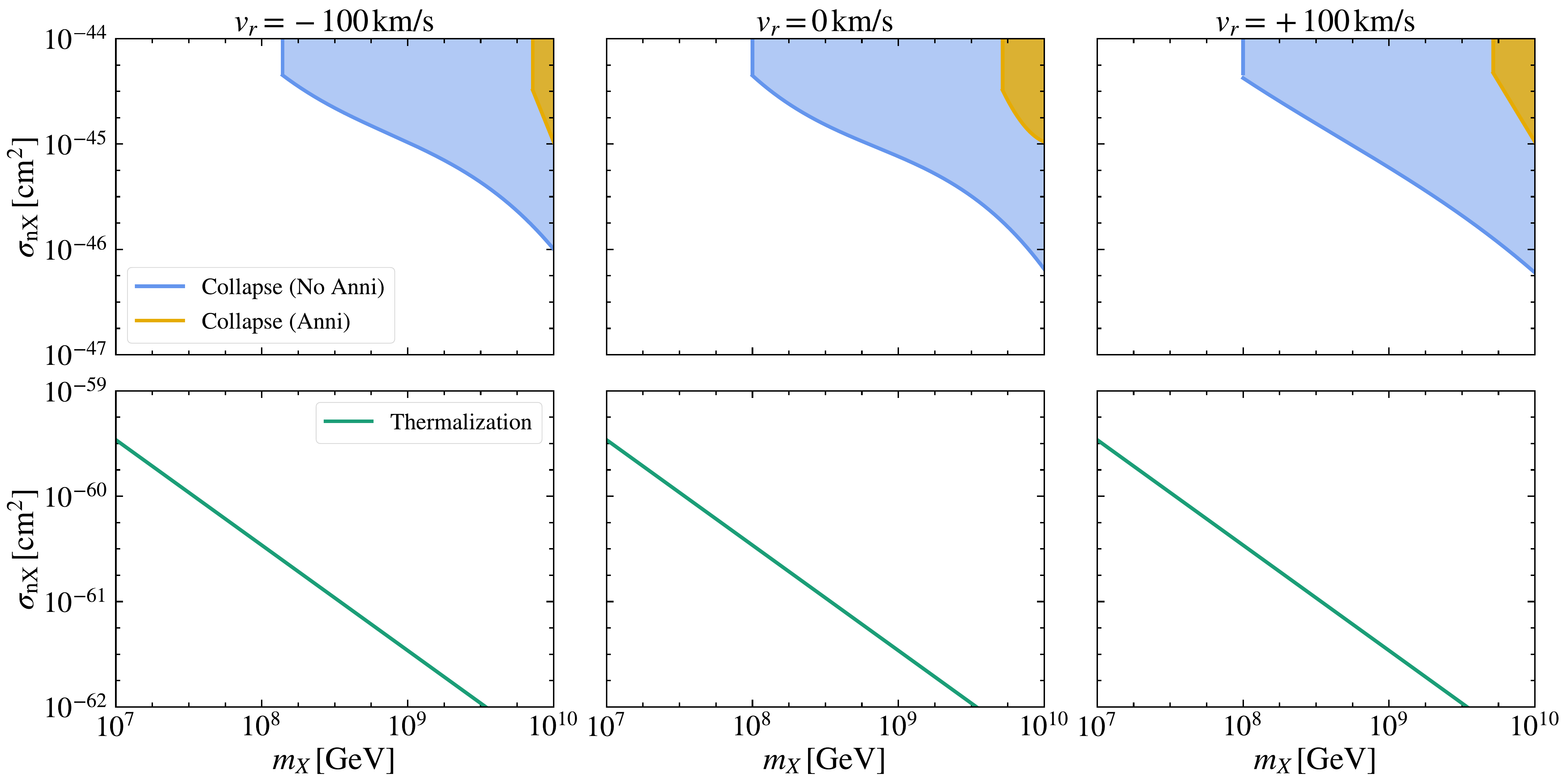} 
        \caption{Constraints on the DM-nucleon cross section $\sigma_{\mathrm{nX}}$ from
        PSR J2129$-$5721. The areas above the orange and blue lines
        satisfy the Collapse Condition with and without DM
        self-annihilation, respectively \textit{(top)}. 
        The areas above the green lines satisfy the Thermalization Condition \textit{(bottom)}. The Growth Condition is fulfilled by
        constraining $m_X \lesssim 10^{10}\,\mathrm{GeV}$. Since the green lines lie far below the blue and orange lines, the feasible regions for NS destruction are actually constrained by the Collapse Condition.
        Overall, the blue and orange
        shaded regions represent the parameter space where neutron star destruction is possible for non-annihilating and annihilating dark matter, respectively.}
        \label{fig:4}
\end{figure}

Considering the fact that NSs exist, i.e. they are not destroyed by the above
process, we can exclude the parameter space where all the three conditions are
met, in which case the NS should have been destroyed by the DM and ended up as a
BH.  Here, we take PSR J2129$-$5721 as an example to show the results in
different scenarios in Fig.~\ref{fig:4}.  We use lines with different colors to
denote the conditions and the shaded areas represent the excluded parameter
space.

We see that in the case where annihilation is considered, DM particles are more
difficult to accumulate, which means that a larger $\sigma_{\mathrm{nX}}$ is needed for
achieving the same value of $N_X$ as in the case of non-annihilation. 
Meanwhile, $N_{\mathrm {coll}}$ only depends on $m_X$, so that the Collapse Condition
is more stringent in the annihilating scenario and it excludes less region. 

\begin{figure}[t] 
            \centering
            \includegraphics[width=\linewidth]{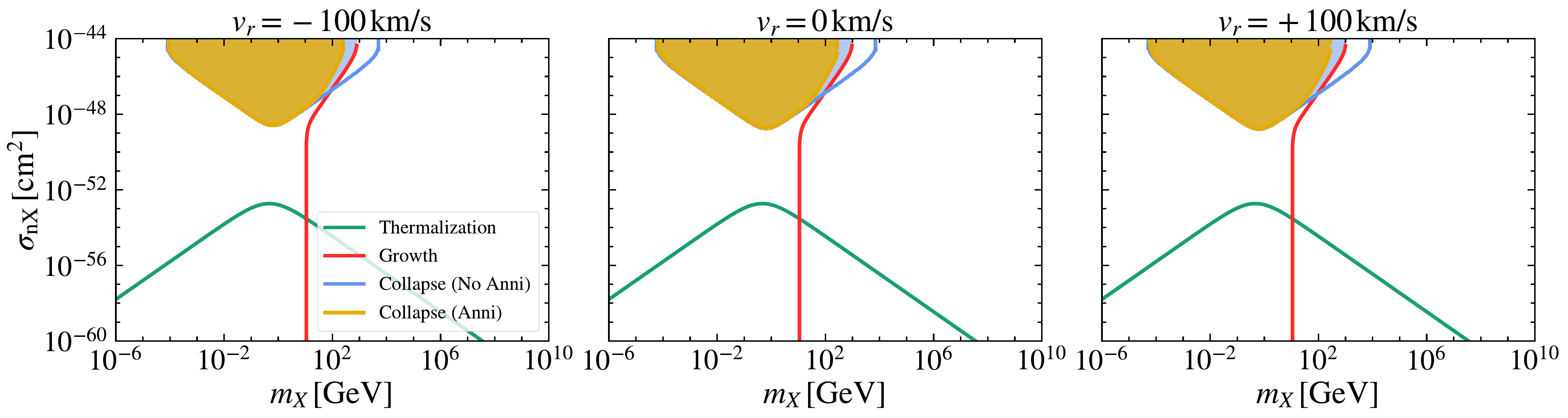} 
        \caption{Constraint on the DM-nucleon cross section $\sigma_{nX}$ from
        PSR J2129$-$5721 for bosonic DM particles. Same labels are used as in
        Fig.~\ref{fig:4}, except that the red line now represents the Growth
        Condition.  The blue and orange shaded regions represent the feasible
        regions where the star can be destroyed for the cases of non-annihilation
        and annihilation, respectively.}
        \label{fig:5}
\end{figure}

To compare with the results in the previous study \cite{Liang:2023nvo}, we also
give results for bosons whose Collapse Condition  is 
\begin{equation}
N_X>N_{\mathrm{BEC}}+N_{\mathrm{chan, bos}} \, \text{,}
\label{eq:32}
\end{equation}
if Bose-Einstein condensation state is formed \cite{Jamison:2013yya}, and we
have
\begin{equation}
\begin{aligned}
N_{\mathrm{BEC}}&=1.2 \times \left(\frac{k_bT_{\mathrm{NS}}}{\sqrt{4\pi G
(\rho_n+3P_n)/3}}\right)^{3} \,,
\end{aligned}
\label{eq:33}
\end{equation}
and $N_{\mathrm{chan, bos}} \sim {m_{\mathrm{pl}}^2}/{m_X^2}$.
Since we do not include a repulsive self-interaction as in
Ref.~\cite{Liang:2023nvo}, we expect a smaller Chandrasekhar limit and a less
excluded area (see Fig.4 in Ref.~\cite{Liang:2023nvo}).  The  Growth Condition
is \cite{Garani:2018kkd}
\begin{equation}
\frac{{\mathrm d} M_{\mathrm{BH}}}{{\mathrm d} t}=\frac{4\pi\lambda
G^2M^2\rho_n}{v_s^3}-\frac{1}{15360\pi G^2M^2}+F_{\mathrm{c}}m_X \,\text{.}
\label{eq:34}
\end{equation}
We give results from PSR J2129$-$5721 for bosonic DM for comparison in
Fig.~\ref{fig:5}.  
As we can see, similar to the effect of self-annihilation,
the fermionic degenerate pressure makes it harder to collapse. 
This results in a much smaller excluded parameter space for fermions compared to bosons.
The difference in the mass range of excluded regions between fermions and bosons is quite large. It also originates from the difference in the collapse condition, i.e. the Chandrasekhar limit.

\begin{figure}[t]
    \captionsetup[subfigure]{singlelinecheck=false, justification=centering} 
    \centering
    \begin{subfigure}{\textwidth} 
            \centering
            \includegraphics[width=1\linewidth]{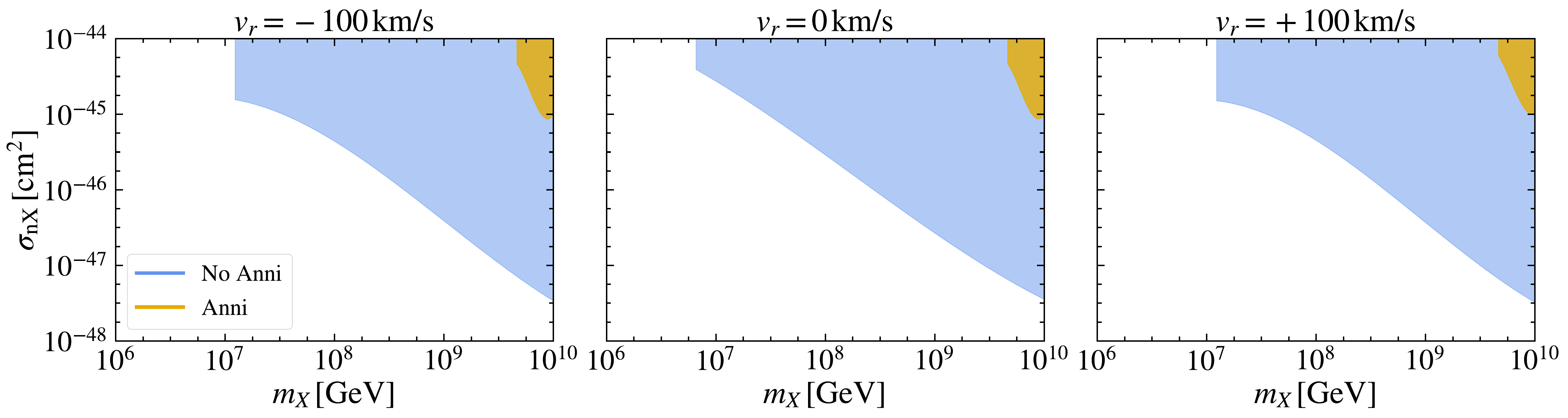} 
        \caption{fermions}
        \label{fig:subim1}
    \end{subfigure}
    \hspace{0.06\textwidth}  
    \begin{subfigure}{\textwidth} 
            \centering
            \includegraphics[width=1\linewidth]{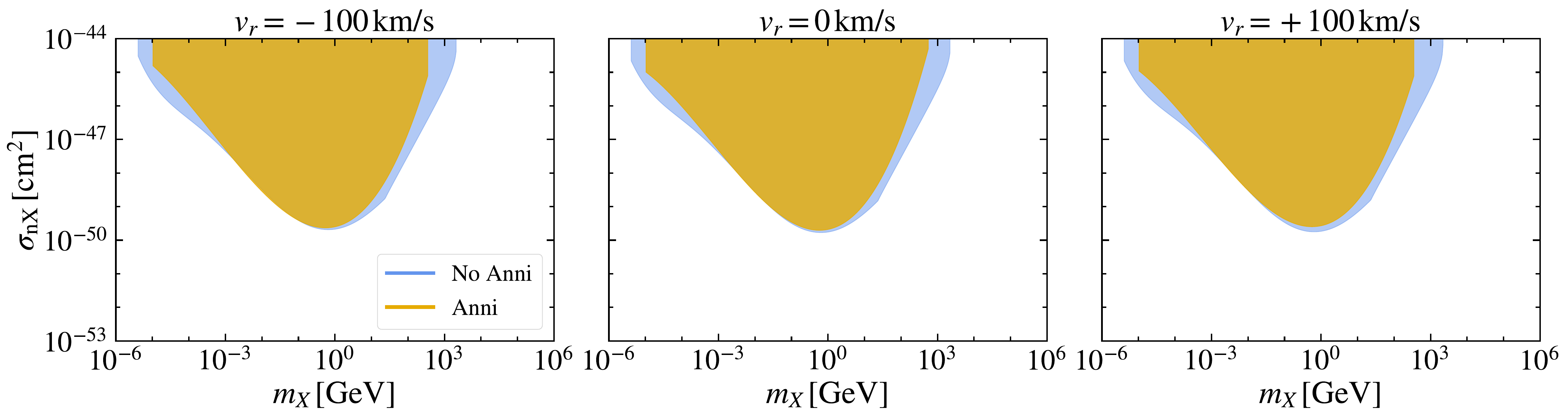} 
        \caption{bosons}
        \label{fig:6(b)}
    \end{subfigure}
    \vspace{0.04cm}  
    
    \caption{Constraints from multiple NSs combined. Same labels are used as in
    Fig.~\ref{fig:4} and Fig.~\ref{fig:5}. \textit{Top:} fermions;
    \textit{bottom:} bosons.}
    \label{fig:6}
\end{figure}

Finally, in the spirit of Liang and Shao~\cite{Liang:2023nvo}, we first consider 413 pulsars in the Milky Way and then apply the Collapse Condition for fermions without annihilation to identify the 12 most constraining ones as our final samples. Combining their results, we obtain the overall constraint presented in Fig.~\ref{fig:6}.

\section{Summary}
\label{sec:4}

We investigated the potential formation of a BH from fermionic DM inside a NS
and the subsequent destruction of the host star, for the first time taking the
effects of the NS motion in the Milky Way into account for fermionic DM. The
effects from DM self-annihilation are also discussed.  From the processes of DM
capture, thermalization, collapse, and BH growth, we have three key conditions
for DM to consume the NS.  By numerically solving the accumulation rate equation
in different scenarios, we identified parameter regions where NS destruction
becomes possible.  Our findings suggest that DM particles are more difficult to
accumulate in scenarios involving self-annihilation, resulting in less stringent
constraints on the DM-nucleon scattering cross-section.  Additionally, we
compared fermionic and bosonic DM in such a scenario to complement the previous
study \cite{Liang:2023nvo}.  This work offers a comprehensive approach to
constraining DM properties through NS observations, contributing to the ongoing
understanding of the nature and interactions of DM.

\section*{Acknowledgments}
We thank the anonymous referees for the valuable comments, and Hong-bo Li for the helpful discussion.
This work was supported by the National SKA Program of China (2020SKA0120300),
the National Natural Science Foundation of China (12405065, 12465013, 12573042), the
Beijing Natural Science Foundation (1242018), the Max Planck Partner Group
Program funded by the Max Planck Society, and the High-performance Computing
Platform of Peking University. 

\appendix

\section{Other possible DM profiles}
\label{sec:5}

\begin{figure}[t]
    \captionsetup[subfigure]{singlelinecheck=false, justification=centering} 
    \centering
    \begin{subfigure}{\textwidth} 
            \centering
            \includegraphics[width=\linewidth]{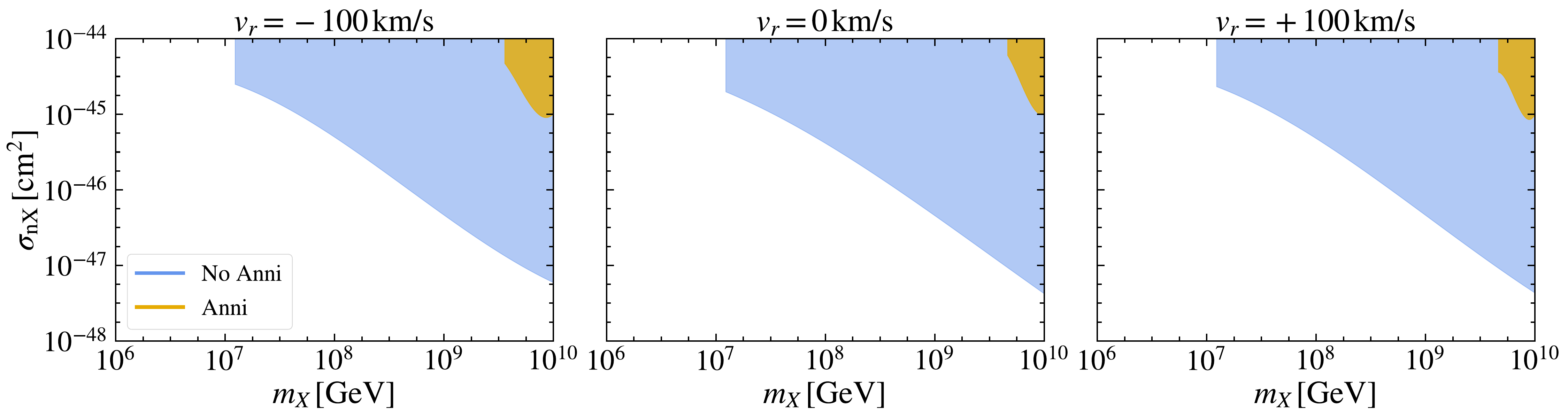} 
        \caption{$\gamma=0.75$}
        \label{fig:subima1}
    \end{subfigure}
    \hspace{0.06\textwidth}  
    \begin{subfigure}{\textwidth} 
            \centering
            \includegraphics[width=\linewidth]{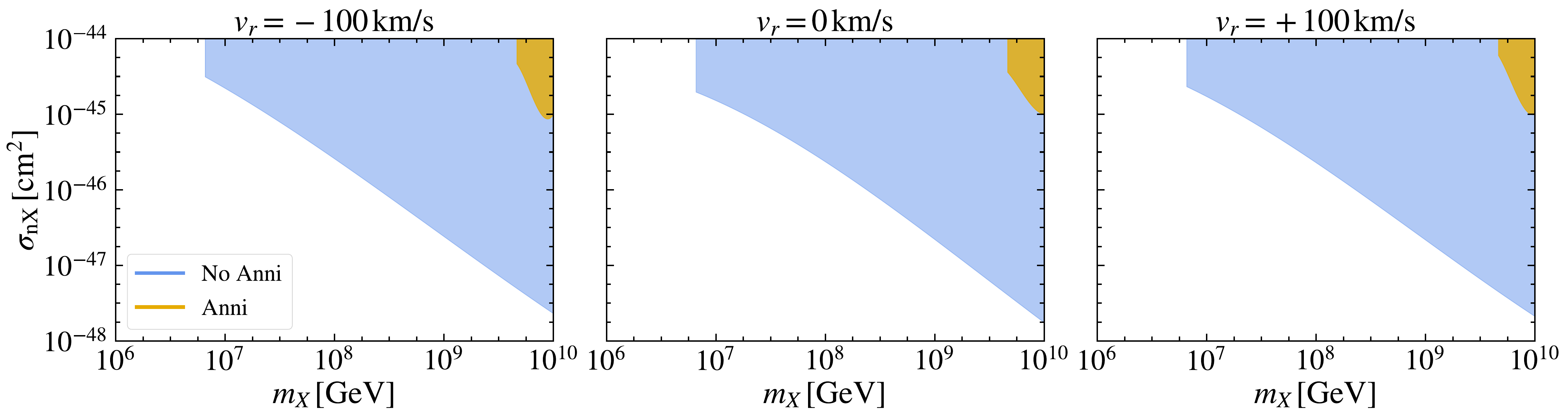} 
        \caption{$\gamma=1.25$}
        \label{fig:subima2}
    \end{subfigure}
    \vspace{0.04cm}  
    
    \caption{Constraint on the DM-nucleon cross section $\sigma_{nX}$ from
    multiple NSs with different $\gamma$. Same labels are used as in
    Fig.~\ref{fig:4}. \textit{Top:} $\gamma=0.75$; \textit{bottom:}
    $\gamma=1.25$.}
    \label{fig:appendix}
\end{figure}

DM distribution in the Milky Way is undetermined. For completeness, we consider
possibilities beyond the NFW profile by changing the value of $\gamma$ while keeping
$r_0=18.6\,\mathrm{kpc}$ and
$\rho_X(r=8.2\,\mathrm{kpc})=0.38\,\mathrm{GeV/cm^3}$ in Eq.~\eqref{eq:5}. In
Fig.~\ref{fig:appendix}, we give the results when $\gamma=0.75$ and $1.25$.

\end{document}